\documentclass[aps,prl,notitlepage,onecolumn,11pt,nofootinbib]{revtex4-1}
\pdfoutput=1

\usepackage{amsmath}
\usepackage{amssymb}
\usepackage{graphicx,subcaption}
\usepackage{braket}

\newcommand\be{\begin{equation}}
\newcommand\ba{\begin{eqnarray}}
\newcommand\ee{\end{equation}}
\newcommand\ea{\end{eqnarray}}

\newcommand{\kk}{\mathbf{k}}
\newcommand{\xx}{\mathbf{x}}

\newcommand{\dd}[1]{\mathrm{d}#1\,}

\begin{document}
\title{Chirality and Circular Polarization in Models of Inflation}
\author{Stephon Alexander}
\affiliation{Department of Physics, Brown University, Providence, R.I. 20912}
\author{Sam Cormack, Robert Sims}
\affiliation{Department of Physics and Astronomy, Dartmouth College, Hanover, NH 03755}

\begin{abstract}
We investigate the possibility that a chiral asymmetry during inflation can manifest as net circular polarization in photons.  Using an example known to produce a helicity imbalance in fermions, we show that superhorizon photon modes produced during inflation acquire net circular polarization.  Modes that reenter the horizon around last scattering can thermalize into the Cosmic Microwave Background while retaining a portion of their net circular polarization.  We also consider the possibility of direct detection of the circular polarization in the CMB.
\end{abstract}

\date{\today}

\maketitle

\section{Introduction}

One of the cornerstone predictions of the physics governing the Cosmic Microwave Background is the polarization of photons.  The polarization spectra gives insight into a handful of important physical processes in the CMB and probes of the early universe.  For example, a detection of B-mode polarization is expected to constrain the gravitational wave amplitude produced during inflation.  The E-mode polarization has been detected and is in agreement with the standard theory of polarization production from a quadrupole anisotropy in the presence of Thompson scattering.  In this work we will demonstrate that it is possible to probe new physics during inflation with circular polarization.  While it is unlikely to produce circularly polarized photons (V-modes) during the recombination epoch, we will show that certain models of inflation can produce V-modes if the inflaton interacts chirally with the photons and electrons.  Such a mechanism of V-mode production during inflation can give a probe into the physics of reheating and baryogengesis.

It has been shown \cite{Alexander,Giovannini:2009ru,Bavarsad:2009hm,Sawyer:2012gn,De:2014qza,Finelli:2008jv} that V-mode polarization can be produced through the presence of anomalous electromagnetic current interactions, magnetic fields in the plasma at the surface of last scattering, and various other mechanisms.  Nonetheless it is difficult to produce a significant amount of V-modes during the recombination epoch.  One loophole is to realize that inflation and subsequently reheating are responsible for producing the CMB, so it is possible that there are operators in inflation that can source circularly polarized photons which persist until the epoch of recombination.  While this idea is general and can arise from gauge models of inflation, for concreteness we will analyze this mechanism in the context of natural inflationary models \cite{Freese:1990rb,Alexander2,Sorbo,Shahin,Adshead}.

Inflation can amplify massless degree of freedom from its vacuum, such as gauge and fermion fields.  Chiral coupling between the inflaton field and these massless fields, such as electrons, could produce a net circularly polarized photon spectrum during inflation.  To provide a concrete example, we investigate the model proposed by Adshead and Sfakianakis \cite{Adshead:2015kza}.  In this model, a chiral coupling between fermions and the axion that drives inflation produces an asymmetry in the handedness of the fermions during inflation \cite{Adshead:2015kza}.  Using the standard electromagnetic coupling to fermions, we investigate the possibility of chiral fermion asymmetry manifesting as V-modes in the CMB.  While the source and magnitude of parity violation is model dependent, once the V-modes are produced, standard electromagnetic interactions should dominate its evolution until last scattering.


\section{Model}
We consider an axion inflation theory with a four-component spinor fermion $\psi$ coupled to gauge field $A^{\mu}$.  There is an additional chiral coupling between the fermion axial current, $J^{\mu 5} = \bar{\psi}\gamma^\mu \gamma^5 \psi$, and the axion.  We work with the following action:
\begin{equation}
\label{model}
S = \int \dd{^4x} \sqrt{-g}\bigg[\frac{M_p^2}{2}R-\frac{1}{4}F_{\mu\nu}F^{\mu\nu} - \frac{1}{2}(\partial\phi)^2 -V(\phi) +\bar{\psi}\left(i\gamma^{\mu}D_{\mu} - m\right)\psi + A_\mu J^\mu+ \frac{C}{f}\partial_\mu \phi J^{\mu 5}\bigg]
\end{equation}
where $C$ is a dimensionless coupling constant, and $f$ is the energy scale.  The covariant derivative for the fermions is given by $D_{\mu}=\partial_{\mu}+\frac{1}{8}\omega_{\mu}^{ab}\left[\gamma_a,\gamma_b\right]$ with spin connection $\omega_{\mu}^{ab}$.  The axion potential $V(\phi)$ is chosen to satisfy the slow-roll conditions for a period of inflation.

To keep contact with \cite{Adshead:2015kza}, we decompose $\psi$ into two-component spinors $\varphi$ and $\eta$ as
\begin{equation}
\psi = \left(\begin{array}{c}
\varphi \\
\eta^\dagger
\end{array}
 \right)
\end{equation}
Writing the fermion current $J^\mu$, the chiral current $J^{\mu 5}$, and the kinetic term for the fermions in terms of the two-spinors,
\begin{align*}
J^\mu &= g(\varphi^\dagger\bar{\sigma}^\mu\varphi - \eta^\dagger\bar{\sigma}^\mu\eta) \\
J^{\mu 5} &= \varphi^\dagger \bar{\sigma}^\mu\varphi+\eta^\dagger\bar{\sigma}^\mu\eta \\
\bar{\psi}\left(i\gamma^{\mu}D_{\mu} - m\right)\psi &= i\varphi^\dagger\bar{\sigma}^\mu D_\mu\varphi+i\eta^\dagger\bar{\sigma}^\mu D_\mu \eta -m(\varphi\eta+\varphi^\dagger\eta^\dagger)
\end{align*}
with $g$ the gauge coupling constant.  We consider a perfect deSitter spacetime with metric
\begin{equation}
ds^2 = a^2(\tau)\left(-d\tau^2 + \delta_{ij}dx^idx^j\right)
\end{equation}
where $\tau$ is the conformal time, $H$ is the comoving Hubble parameter and assumed to be constant.  The scale factor in conformal time is given by $a(\tau) = \frac{-1}{H\tau}$.

The equation of motion for the gauge field can be found as
\begin{gather}
{A^0}'' +{6\mathcal{H}A^0}' -\nabla^2 A^0 +12\mathcal{H}^2 A^0 = a^2 g(\varphi^\dagger\bar{\sigma}^0\varphi - \eta^\dagger\bar{\sigma}^0\eta)\\
A_i''-\nabla^2 A_i -2a^2\mathcal{H}\partial_i A^0= a^2 g(\varphi^\dagger\bar{\sigma}_i\varphi - \eta^\dagger\bar{\sigma}_i\eta)
\end{gather}
where a prime denotes a conformal time derivative and we define the conformal Hubble parameter as $\mathcal{H} = \frac{a'}{a}$.  The presence of sources does not allow us to use the Coulomb gauge and set the $A^0$ term to zero.  However, the $A^0$ term in the equation of motion for the spatial components of $A^\mu$ will act as correction to the current.  For small charge densities, this term can be ignored.  Furthermore, as we will see, the equation for $A^0$ will not need to be solved in order to find the amount of circular polarization.  To make direct connection with \cite{Adshead:2015kza}, we rescale the fermion field as $\Psi \rightarrow a^{3/2}\Psi$.  Looking at the equation of motion of the Fourier modes,
\begin{equation}
A_{ik}'' +k^2 A_{ik} = a^{-1}J_{ik}. \label{rsGaugeEoM}
\end{equation}
We have defined the Fourier transformed current as
\begin{equation}
J^\mu_k = g\int\dd{^3x}e^{-i\mathbf{k}\cdot\mathbf{x}}(\varphi^\dagger\bar{\sigma}^\mu\varphi - \eta^\dagger\bar{\sigma}^\mu\eta). \label{FTCurrent}
\end{equation}
The fermion two-spinors can be expanded in terms of creation operators as
\begin{gather}
\varphi_\alpha(\xx,\tau)=\sum_\lambda \int\frac{\dd{^3k}}{(2\pi)^3}\left[ x^\lambda_\alpha(\kk,\tau)b^\lambda_ke^{i\kk\cdot\xx}+y^\lambda_\alpha(\kk,\tau)\bar{b}^{\dagger\lambda}_k e^{-i\kk\cdot\xx}\right]\label{quantphi}\\
\eta_\alpha(\xx,\tau)=\sum_\lambda \int\frac{\dd{^3k}}{(2\pi)^3}\left[ x^\lambda_\alpha(\kk,\tau)\bar{b}^\lambda_ke^{i\kk\cdot\xx}+y^\lambda_\alpha(\kk,\tau)b^{\dagger\lambda}_k e^{-i\kk\cdot\xx}\right]\label{quanteta}
\end{gather}
where $\alpha$ explicitly labels the two spinor indices, and the index $\lambda$ labels the two independent fermion states, which we can take to be helicity eigenstates.  Adshead and Sfakianakis \cite{Adshead:2015kza} explicitly expand the spinors $x^\lambda$ and $y^\lambda$ into a time dependent ampitude and the helicity eigenspinor.  However, for the following analysis, we need only refer to the results of their work.

\section{Calculation of V-Modes}
In order to calculate the amount of circular polarization produced, we must find an exact solution to equation (\ref{rsGaugeEoM}).  The general solution is given by
\begin{equation}
\vec{A}_k (\tau) = \vec{\b{A}}_{k}(\tau) + i\int \dd{\eta}a^{-1}(\eta)G_{k}(\tau,\eta)\vec{J}_k(\eta) \label{GenSol}
\end{equation}
where $\vec{\b{A}}_{k}$ is the background solution for the homogeneous form of equation (\ref{rsGaugeEoM}) and $G_{k}(\tau,\eta)$ is the Green's function.  Explicit forms of $\vec{\b{A}}_{k}, G_{k}$, as well as details of the calculation, can be found in the Appendix.

The Stokes parameter V can be calculated, in the circular polarization basis $\epsilon^+_{\mu}, \epsilon^-_{\mu}$, as the difference in intensities of the two polarization
\begin{equation}
V(\hat{n}) = |\epsilon^{+*}_{\mu}(\hat{n})E^{\mu}|^2 - |\epsilon^{-*}_{\mu}(\hat{n})E^{\mu}|^2 \label{defV}
\end{equation}
where $E^{\mu} = F^{\mu\nu}u_{\nu}$.  For $u_{\mu} = (a,0,0,0)$, we have the general form for the electric field $E_{\mu} = a^{-2}\left(\partial_{\mu}A_0 - \partial_0 A_{\mu}\right)$.  However, when the transverse polarization tensor is contracted with the electric field, the terrm $\partial_\mu A_0$, which is proportional to $k_\mu A_0$, will vanish.  Then $|\epsilon_{\lambda}^{\mu *}(\hat{n})E_{\mu}|^2 = |a^{-2}\epsilon_{\lambda}^{\mu *}(\hat{n})\frac{\partial}{\partial\tau}A_{\mu}|^2$.

Since $A^{\mu}$ is a quantized field, to calculate the intensity of a particular polarization we have taken expectation values of $|\epsilon_{\lambda}^{\mu *}E_{\mu}|^2$ on the vacuum.  Using the general solution in equation (\ref{GenSol}), the V parameter can be calculated from equation (\ref{defV}) as
\begin{align}
V = a^{-4}(\tau)\int\frac{\dd{^3k}\dd{^3k'}\dd{\eta}\dd{\eta'}}{(2\pi)^6}&e^{i(k-k')x}a(\eta)a(\eta')\left[\frac{\partial}{\partial\tau}G_k(\tau,\eta)\frac{\partial}{\partial\tau}G^{*}_k(\tau,\eta')\right]\nonumber\\
&\times \Bra{0} J_i(k,\eta)J^{*}_j(k',\eta')\Ket{0} \left[\epsilon_i^{+*} \epsilon_j^+ - \epsilon_i^{-*} \epsilon_j^-\right]
\end{align}
Using equations (\ref{FTCurrent}-\ref{quanteta}), the V parameter can then be written in terms of the fermion functions $x,y$ as
\begin{align}
&V = \frac{4g^2}{a^4(\tau)}\sum_{\alpha,\beta}\int\frac{\dd{^3k}\dd{\eta}\dd{\eta'}}{(2\pi)^6}a(\eta)a(\eta')\left[\frac{\partial}{\partial\tau}G_k(\tau,\eta)\frac{\partial}{\partial\tau}G^{*}_k(\tau,\eta')\right]\text{Im}\left[\epsilon^+_i \epsilon^{+*}_j\right]\nonumber \\
& \times i\int\dd{^3p}\Big\{y^{\alpha\dagger}(\textbf{p},\eta)\bar{\sigma}_{j}x^{\beta}(-\textbf{p}-\textbf{k},\eta) \left[x^{\beta\dagger}(-\textbf{p}-\textbf{k},\eta')\bar{\sigma}_{i}y^{\alpha}(\textbf{p},\eta')-x^{\alpha\dagger}(\textbf{p},\eta')\bar{\sigma}_{i}y^{\beta}(-\textbf{p}-\textbf{k},\eta')\right]\Big\}. \label{VCalc}
\end{align}
The sum of $\alpha, \beta$ is over the helicities states $\{+,-\}$ of the fermions and there is an implicit sum over spatial coordinates $i,j$.  In this calculation, the intensity of the background field, $E^\mu_0$, is independent of polarization.  This recovers the fact that an absence of current leads to no net circular polarization.  Further, the ``cross'' terms containing factors of the form $\Bra{0}E^\mu_{k,0}J^{\nu}_{k}\Ket{0}$ are zero. Because the Hilbert space of the entire system can be written as the tensor product of Hilbert spaces for the gauge field and fermions, the cross terms become proportional to the VEV of the electric field.

In the limit where the $C\rightarrow 0$, there is no fermion helicity asymmetry, and the amplitudes of the $p$-integral become symmetric under the exchange of $i$ and $j$.  Since $\text{Im}\left[\epsilon^+_i \epsilon^{+*}_j\right]$ is antisymmetric under this exchange, once again, we recover the limit that $V=0$.  However, since the amplitudes of the fermion helicity eigenstates are not equivalent \cite{Adshead:2015kza}, we find nonzero V-modes.

When interested in only the superhorizon modes, we can take the limit of V when $\left|k\tau\right|\ll 1$.  The only remaining $\tau$ dependence comes from the factor of the derivatives of the Green's function.  From the Appendix, the retarded Green's function has the form
\begin{equation}
G_k(\tau,\eta) = -\frac{i}{k}\theta(\tau-\eta)\sin k(\tau-\eta).
\end{equation}
Imposing the step function in the Green's function reduces the integral in equation (\ref{VCalc}) to include only the region where $k\eta = y < x = k\tau$.  Then, the Stoke's parameter becomes a product of two physical functions.
\begin{equation}
 V(\tau) = 4g^2H^4\tau^4\int\frac{\dd{^3k}\dd{\eta}\dd{\eta'}}{(2\pi)^3}\cal{I}\rm_{1}(k,\eta,\eta')\cal{I}\rm_{2}(k,\eta,\eta')
\end{equation}
 
These two integrals contain useful physical information about the evolution of the circular polarization.  We see that the function 
\begin{equation}
\cal{I}\rm_{1}(k,\eta,\eta')=a(\eta)a(\eta')\cos\left(k(\tau-\eta)\right)\cos\left(k(\tau-\eta')\right)\text{Im}\left[\epsilon^+_i\epsilon^{+*}_j\right]
\end{equation}
projects the modes produced by fermions into physical outgoing photon states.  Also, the second integral, $\cal{I}\rm_{2}(k,\eta)$ illustrates how the chirality of the electrons get transferred to the photon's circular polarization since 
\begin{align} 
\cal{I}\rm_{2}(k,\eta,\eta') =i\sum_{\alpha,\beta}\int\frac{\dd{^3p}}{(2\pi)^3}y^{\alpha\dagger}(\textbf{p},\eta)\bar{\sigma}^{j}x^{\beta}(-\textbf{p}-\textbf{k},\eta) &\big[x^{\beta\dagger}(-\textbf{p}-\textbf{k},\eta')\bar{\sigma}^{i}y^{\alpha}( \textbf{p},\eta')\nonumber\\
&-x^{\alpha\dagger}(\textbf{p},\eta')\bar{\sigma}^{i}y^{\beta}(-\textbf{p}-\textbf{k},\eta')\big]
\end{align}
involves an anti-symmetric product of the electron's helical mode-functions, encoding the net-circular polarization.

As comoving time progresses, the conformal time $\tau$ goes to zero, therefore at early times $\tau$ increases and as seen in the above equation the intensity for $V$ will likewise get enhanced by a factor of $\tau^{4}$.  This is because the Stokes parameters are related to the intensity of light, which diminishes as the universe expands.  Since the scale factor $a(\tau)$ is inversely proportional to $\tau$, we see that this intensity is scaling as $a^{-4}$ as we expect for radiation.  Because of the complexity of the fermion mode functions a numerical treatment is needed to analyze the powerspectra of $V$.  The main purpose of this work is to demonstrate that V can be produced during inflation and can in principle be non-vanishing in the CMB.  We will report the results of a numerical treatment of V-mode powerspetrum in an upcoming work \cite{AMS}.



\section{Possibility of Detection}
We can find the wavenumber of photons just re-entering the horizon as $k = aH$.  The Hubble parameter at a given cosmological redshift, z, is
\begin{equation}
H^2(z) = H^2_0\left(\Omega_\Lambda+\Omega_M(1+z)^3 + \Omega_R(1+z)^4\right)
\end{equation}
where $\Omega_\Lambda, \Omega_M, \Omega_R$ are the density parameters for dark energy, matter, and radiation, respectively, and $H_0$ is the Hubble constant (about $70 \text{ km s}^{-1} \text{Mpc}^{-1})$.  Then, at last scattering, the wavenumber of photons just re-entering the horizon is roughly $k \sim 20H_0$.  If we are only searching for V-mode polarization in photons that are just sub-Hubble at last scattering, the wavelengths are well beyond what is currently detectable.  Therefore, in order to directly detect the presence of the circular polarization, we must look to thermal photons present in the CMB.   As usual, we expect that these superhorizon photons will thermalze via the appropriate reheating mechanism.  In this case the inflation pseudoscalar field can undergo coherent oscillations and will thermalize the gauge fields \cite{Evan,Amin:2014eta}.\footnote{The photons may also thermalize due to their interaction with the electron during the electroweak phase transition.  We thank Alan Guth for this suggestion which we shall pursue in a forthcoming paper.}   

Compton scattering between electrons and circularly polarized photons is one of the primary methods in which the photons will thermalize.  This Compton scattering will also result in spreading the net helicity over all degrees of freedom, resulting in a net helicity in both the population of electrons and photons.  Since the electrons are massive, there is mixing between right and left handed states which will move into equilibrium, removing any net helicity in the electrons.  As a result, the process of Compton scattering will couple the net helicity in photons to a helicity sink.  For V-modes to be detectable in the CMB, general V-modes must survive through standard electromagnetic interactions after electroweak symmetry breaking.

In considering general Compton scattering, the collision term for the Boltzmann equation can be written as
\begin{equation}
C[f(k)] = \int \mathcal{D}p\mathcal{D}k'\mathcal{D}p' (2\pi)^4\delta^{(4)}(p'+k'-p-k)|\mathcal{M}|^2\left[f(p')f(k')- f(p)f(k)\right] \nonumber
\end{equation}
where $\mathcal{D}p = \frac{d{^3p}}{(2\pi)^3}\frac{1}{2E}$ and $\mathcal{M}$ is the matrix element for the scattering event $k + p \rightarrow k'+p'$.  This collision term dictates how the distribution function $f(k)$ will change over time.  We can recover the collision term associated with the change in number density of particle $k$ by integrating $\mathcal{D}k$.

In particular, we consider the helicity flipping interaction $e_R\gamma_L \leftrightharpoons e_L\gamma_R$ and consider the collision term for $\gamma_L$.  The high energy case, $\mathcal{M}$ has the form \cite{PeskinSchroeder}
\begin{equation}
\mathcal{M}(e^{-}_{R}\gamma_{L}\rightarrow e^{-}_{L}\gamma_{R}) \approx \frac{4e^2 m/\omega}{\chi^2+m^2/\omega^2} \label{helflipM}
\end{equation}
where $\chi=\pi-\theta$, $\theta$ is the scattering angle, and $\omega$ is the center of mass energy of the photon.  By definition, the number density of a population is
\begin{equation}
n(t) = \frac{g}{(2\pi)^3}\int d{^3p}f(p) \nonumber
\end{equation}
where $g$ is the degeneracy factor of the species.  If we consider the distribution function for both right and left handed photons then $f(p_L)=f(p_R)$, so we have equal number densities $n_L, n_R$.  Thus, we must also scale the photon distribution functions by $n_{L,R}/(n_L+n_R)$ to recover the proper number densities. This effectively makes the degeneracy factor of all photons a weighted average of the two helicity states.  Using Maxwell-Boltzmann statistics for all particles and imposing energy conservation, the Boltzmann equation can be written as
\begin{equation}
\dot{n}_{L,\gamma} + 3Hn_{L,\gamma} = \int \mathcal{D}k\mathcal{D}p\mathcal{D}k'\mathcal{D}p' (2\pi)^4\delta^{(4)}(p'+k'-p-k)|\mathcal{M}|^2e^{-E_{tot}/k_{B}T}\left(\frac{n_{R,\gamma}z_L-n_{L,\gamma}z_R}{n_{L,\gamma}+n_{R,\gamma}}\right) \nonumber
\end{equation}
where $z_i$ is the fugacity of the right/left handed electrons, $z_i = e^{\mu_i/k_BT}$. Since we consider Maxwell-Boltzmann distributions for all particles, we assume thermal equilibrium between photons and electrons before electroweak symmetry breaking.  The helicity flipping Compton scattering is only allowed after the electroweak phase transition, with a constant number of electrons.  Since the photons remain in thermal equilibrium during the helicity flipping process, the quantity in parentheses is only time dependent.  The remaining integral is some constant related to the kinematics of the process, which we will define as $\alpha$. \footnote{Explicitly, $\alpha = \int \mathcal{D}k\mathcal{D}p\mathcal{D}k'\mathcal{D}p' (2\pi)^4\delta^{(4)}(p'+k'-p-k)|\mathcal{M}|^2e^{-E_{tot}/k_{B}T}$.}  Then, we can rewrite the left side of the Boltzmann equation as
\begin{equation}
\dot{n}_{L,\gamma} + 3Hn_{L,\gamma} = a^{-3}\frac{\partial}{\partial t}(a^3 n_{L,\gamma})
\end{equation}
Compton scattering conserves the total photon number, hence the number density is proportional to $a^{-3}$.  Subtracting the equations for right and left handed photons, we find
\begin{equation}
a^{-3}\frac{\partial}{\partial t}\left[a^3\left(n_{L,\gamma}-n_{R,\gamma}\right)\right] = -\frac{2\alpha z_e}{n_{\gamma}}(n_{L,\gamma}-n_{R,\gamma})
\end{equation}
This equation can be solved exactly for the difference in helicities as a function of time as
\begin{equation}
\left(n_{L,\gamma}-n_{R,\gamma}\right)(t) \propto a^{-3}\exp\left[-\int_{t_0}^{t}\lambda a^3(t') d{t'}\right] \label{numdensol}
\end{equation}
where $\lambda = \frac{2\alpha z_e}{(n_{\gamma}a^3)|_{t_0}}$, the denominator is the total number of photons at time $t_0$.  If we included interactions so that the total photon number could change, it would remain a dynamic variable in the time integral.  Since the V Stokes parameter is related to the difference in particle number density, once the net circular polarization is allowed to interact through a helicity flipping process, the observed V-modes are exponentially suppressed.  However, the exponential quantity can be calculated between an initial time at electroweak symmetry breaking and final time at last scattering, and is found to be a perturbative effect on the $a^{-3}$ scaling.  The primary reason this exponential is small is because the number density of electrons is much smaller than the number density of photons in the early universe.  Most of the helicity flipping process occurs around electroweak symmetry breaking due to the high interaction rate.

In general, the photons and electrons may not be in thermal equilibrium at electroweak symmetry breaking.  Furthermore, if we allow the superhorizon V-modes to reenter the horizon and interact with the electrons, this will add a nonthermal distribution to the photon population.  For these photons to enter the CMB, they must thermalize in order to reduce the spectral distortions in the CMB.  This thermalization process will also lead to the ``decay'' of circular polarization in these photons.  The timescales associated with thermalization via Compton scattering and the ``decay'' of the photon circular polarization will, in general, be different.  We will label these timescales $t_C$ and $t_D$, respectively.  While these timescales will be investigated in more detail in a future paper, we provide a proof of concept here.

In the following analysis, we consider a single photon interacting with a bath of electrons.  Since there is not a notion of a thermal distribution for a single photon, we treat this thermalization process as a single Compton scattering event.  When considering a distribution of photons, multiple scatterings are required to reach a thermal state, so this single interaction provides a lower bound on the thermal timescale.  Similarly, we consider a single helicity flipping interaction as the timescale for the ``decay'' of the circular polarization.  If we consider the collision term in the Boltzmann equation, we find the scattering rate is $\Gamma = n_e\sigma$ where $n_e$ is the number density of electrons and $\sigma$ is the cross section of the appropriate interaction \cite{PeterUzan}.  Then, we have timescales
\begin{align*}
t_C &= \left(n_e\sigma_C\right)^{-1} \\
t_D &= \left(x_R n_{e,L} \sigma_{C,h} + x_L n_{e,R}\sigma_{C,h}\right)^{-1}
\end{align*}
where $\sigma_C$ is the total Compton cross section, $\sigma_{C,h}$ is the helicity-flipping Compton cross section, and $x_{L,R} = n_{L,R}/n_{\gamma}$ is the ratio of left, right handed photons number density to the total photon number density.  These factors show up because we are considering a single photon to represent the entire population of photons re-entering and thermalizing.

Obviously, the helicity-flipping cross section must be contained in the total Compton cross section, so $\sigma_{C,h} \leq \sigma_{C}$.  Furthermore, we know that $n_{\gamma,L},n_{\gamma,R}\leq n_{\gamma}$ and $n_{e,L},n_{e,R}\leq n_{e}$.  If we assume that all electrons and photons are either right or left circularly polarized, then 
\begin{equation}
n_{L}+n_{R}= n_{total}
\end{equation}
for both photon and electrons.  Therefore, we can approximate the ratio of timescales as
\begin{equation}
\frac{t_C}{t_D} \approx \left[\frac{n_{\gamma}n_{e}\sigma_C}{\left(n_{\gamma,L}n_{e,R}+n_{\gamma,R}n_{e,L}\right)\sigma_{C,h}}\right]^{-1} \leq \frac{\sigma_{C,h}}{\sigma_{C}} \leq 1
\end{equation}
This relation is true regardless of the assumptions made about the populations of electrons and photons.  Therefore, thermalization of the photons will occur before the circular polarization is washed out.  There is a window before last scattering in which photons can re-enter the horizon with enough time to thermalize, making them detectable, but late enough for so that there is still a trace of their initial circular polarization.

For a concrete example, we can look at the matrix elements for high energy Compton scattering. The u-channel matrix elements have been calculated \cite{PeskinSchroeder} in the high energy limit as
\begin{equation}
\mathcal{M}(e^{-}_{R}\gamma_{R}\rightarrow e^{-}_{R}\gamma_{R}) \approx \frac{4e^2\chi}{\chi^2+m^2/\omega^2} \nonumber
\end{equation}
and the helicity flipping matrix element is given by equation (\ref{helflipM}).  After electroweak symmetry breaking, the energy of the photons remains high enough that the photon helicity flipping process is highly suppressed.  However, as the Universe expands, the helicity flipping process becomes more important.  Photons with a net helicity that are allowed to interact soon after electroweak symmetry breaking will have ample time to wash out their helicity. So we only consider photons that reenter the horizon within a single thermalization timestep from last scattering.  At this time, the high energy approximations used to calculate the above matrix elements no longer hold.  However, a similar calculation of the cross sections at low energies can be done to find
\begin{equation}
1-\frac{t_C}{t_D} \approx \left(\frac{\omega}{m}\right)^2. \nonumber
\end{equation}
At last scattering, this gives a ratio of $10^{-12}$.  Therefore, we expect $10^{-12}$ of the initial energy density of the V-modes reentering the horizon during the final thermalization timestep to persist into the CMB.  In an upcoming work, we will provide an explicit preheating mechanism based on our Lagrangian \cite{model}, where the inflation decays into circularly polarized photons and the back reaction on the fermions are taken into account \cite{AMS}.

\section{Consequences of Electron Number}
In the model considered, electrons act as both a source and sink to the photon circular polarization.  During inflation and reheating, the inflaton coupling drives the production of electrons with a preferred handedness, which subsequently produces photons with a net circular polarization.  After electroweak symmetry breaking, the electron mass allows for helicity flipping Compton scattering, which converts the circular into linear polarization.  We now consider two possible histories.

From equation (\ref{GenSol}), we see that the more total electrons produced, the more photons will be produced.  If we only allow for the production of electrons through the inflaton coupling, then this production of electrons will produce a definite amount of V-modes.  Since the ratio of V-modes to total intensity of light is given by the exponential factor in equation (\ref{numdensol}), we need a large amount of electrons present around electroweak symmetry breaking to substantially diminish the V-modes present.  As a result, another mechanism is needed to explain why the number density of electrons drops by several orders of magnitude.  One such effect could be recombination, as this will only reduce the free electron number density after a majority of the helicity flipping process has occurred.

We can avoid the electron number density issue by setting up initial conditions so that the predicted intensity of V-modes are consistent with observations.  First, we see from equations (\ref{defV}, \ref{VCalc}) that, in general, photons of both helicity are produced, so the intensity of the V-modes are strictly less than the total intensity of light.  This effect is amplified during reheating when the inflaton oscillates about the minima of its potential.  The derivative coupling will change sign, overproducing different handed fermions in different phases of oscillation \citep{Adshead:2015kza}.  At one time, left-handed photons may be produced due to the overproduction of left-handed electrons, while at another, right-handed photons will be produced due to the overproduction of right-handed electrons.  This will allow the number of photons to be high while the net circular polarization remains relatively small.  The dynamics of a given inflation and reheating model will largely influence the observed V-modes.  In particular, if the inflaton is weakly coupled to other fields, reheating will be a slow process and the circular polarization in photons will be small compared to the total energy in photons.

\section{Discussion}
In this work we argue that inflationary models where the inflaton field has chiral couplings to electrons the net helicity of amplified electrons will bias the sourcing of circularly polarized superhorizon photon modes.  Provided that these photons are thermalized during reheating, we solved the Boltzmann equations for a general population of circularly polarized photons produced from reheating to recombination and showed that Compton scattering will intraconvert the circularly polarized photons into linearly polarized ones.  However this intraconversion is exponentially suppressed and depends on the initial electron number density and the details of reheating.  Therefore, it is plausible that a non-negligible fraction of the polarized photons in the CMB are circularly polarized.   In an upcoming work, we will provide a detailed analysis of the reheating mechanism where the circularly polarized photons are thermalized by their interactions with the electrons which are also produced during inflation.  This analysis will give us the initial conditions to determine what percentage of the CMB is circularly polarized.

\section{Appendix}
We are interested in finding solutions to the gauge field equation of motion when there is no current.  In particular, we will find the limiting behavior of the general solution and apply these limit to solve for the Green's function.  Explicitly written, we are looking for solutions to the equations
\begin{align}
f''_k(\tau) + k^2f(\tau) &= 0 \label{BackgroundEoM}\\
G''_k(\tau,\eta) + k^2G_k(\tau,\eta) &= -i\delta(\tau-\eta) \label{GreensDif}
\end{align}
where $\delta(\tau-\eta)$ is the Dirac delta function.

The general solution to equation (\ref{BackgroundEoM}) can be written as the linear combination
\begin{equation}
f_k(x) = Ae^{-ix}+Be^{ix}\label{BackgroundSol}
\end{equation}
where $x=k\tau$.  Starting inflation from the Bunch-Davies vacuum at $\tau \rightarrow -\infty$ gives the solution $f_k(\tau) = \frac{1}{\sqrt{2k}}e^{-ik\tau}$.  This solution will act as the background solution for the gauge field.

We can solve for the Green's function as the solution to equation (\ref{GreensDif}).  The Dirac delta splits the $(\tau,\eta)$ plane into two regions.  In each of these regions, the Green's function is continuous and differentiable and solves the background equation (\ref{BackgroundEoM}).  Therefore, each region has general solution given by equation (\ref{BackgroundSol}).  With these conditions, we can write the Green's function as
\[
  G_{k}(x,y) = \left\{\def\arraystretch{1.2}%
  \begin{array}{@{}c@{\quad}l@{}}
    Ae^{-ix}+De^{ix} & x<y\\
    Be^{-ix}+Ce^{ix} & x>y\\
  \end{array}\right.
\]
where $x = k\tau, y = k\eta$.  If we impose the condition that inflation starts from the Bunch-Davies vacuum at $\tau\rightarrow -\infty$, we consider the retarded Green's function.
\begin{equation}
G_k(x,y) = \theta(x-y)\left(Be^{-ix}+Ce^{ix}\right)
\end{equation}

The Green's function should also be continuous across the boundary at $x=y$, and should have a discontinuous derivative.  Explicitly, these conditions give the equations
\begin{gather}
C= -Be^{2iy}\\
\lim_{\epsilon\to 0^{+}}\left[\frac{\partial G_k}{\partial x}\bigg|_{\tau=\eta+\epsilon}-\frac{\partial G_k}{\partial x}\bigg|_{\tau=\eta-\epsilon}\right] = -\frac{i}{k}.
\end{gather}
Under these conditions, the retarded Green's function is
\begin{equation}
G_k(\tau,\eta) = -\frac{i}{k}\theta(\tau-\eta)\sin k(\tau-\eta)
\end{equation}

\section{Acknowledgment}We wish to thank Katie Freese, Brian Keating, David Kaiser and Subodh Patil for reading a draft of this paper and making useful suggestions.  S.A thanks Brown University for financial support of this work.  S.C and R.S thank Dartmouth College for financial support of this work.

\end{document}